\documentclass[aps,prd,groupedaddress]{revtex4-1}
\usepackage{epsfig}
\usepackage{amsfonts,amssymb}
\usepackage{color}

  \usepackage{slashed}
  \usepackage{url}
\usepackage{bm}
\usepackage{verbatim}
\usepackage{float}

\begin{document}

\title{Exact solutions to the sector of the Standard Model Higgs}


\author{Marco Frasca}\email{marcofrasca@mclink.it}
\affiliation{via Erasmo Gattamelata, 3, 00176 Roma, Italy}


\begin{abstract}
Recently, we have found an exact solution to the full set of Dyson-Schwinger equations of the non-interacting part of the Higgs sector of the Standard Model obtained by solving the 1-point correlation function equation. In this work we extend this analysis considering also the other possible solution that is the one experimentally observed in the Standard Model.
Indeed, the same set of Dyson-Schwinger equations can be exactly solved for the Standard Model with a constant as a solution for the 1-point correlation function.
Differently from the Standard Model solution, the one we have found has a mass spectrum of a Kaluza-Klein particle. This could be a clue toward the identification of a further space dimension. Gap equations are obtained in both cases as also the running self-coupling equations. 
\end{abstract}


\maketitle

\section{Introduction}


The Higgs mechanism was discovered in the sixties \cite{Englert:1964et,Higgs:1964ia,Higgs:1964pj,Guralnik:1964eu,Higgs:1966ev,Kibble:1967sv} and successfully introduced in the Standard Model of the elementary particle by Weinberg and Salam  \cite{prl_19_1264,sm_salam}. After several decades of search, the accomplishment of the discovery of the Higgs particle was concretized by the LHC in the 2012  \cite{Aad:2012tfa,Chatrchyan:2012ufa}. 
The non-interacting part of the Higgs sector, that is obtained by removing couplings with fermions and gauge fields, is suspected to be trivial (see \cite{triv} and Refs. therein). Such a question was recently addressed in \cite{Frasca:2006yx,Suslov:2010rk} for the strong coupling limit and from the exact solution of the Dyson-Schwinger equations in \cite{Frasca:2015wva}. This is essentially a mathematical problem that is still considered open.


Quite recently we devised a different solution of the Higgs sector by exactly solving the set of the Dyson-Schwinger equation for the scalar field secor in the Standard Model \cite{Frasca:2015wva}. This result seems to work quite well for a Yang-Mills theory where a constant solution cannot exist being this a gauge theory \cite{Frasca:2015yva}. For the Higgs field, it is exactly the opposite: We have a constant exact solution for the 1-point correlation function. Indeed, all the results coming out from LHC are showing us that this is the preferred choice by nature. So, it is rather interesting to study the set of Dyson-Schwinger equation for this case and the gap equations that are obtained in this way for both the possible solutions.

The spectrum of the solution proposed in \cite{Frasca:2015wva} implies the existence of higher excited states in the Higgs spectrum similar to a Kaluza-Klein particle. Notwithstanding these states are not observed in current LHC experiments, supporting the simple spectrum solution, some other independent theoretical analyses based on the renormalization group back the former possibility \cite{Hill:2019cce}. Currently, ongoing studies at CERN could not completely exclude further resonances of the Higgs particle. The search is pursued yet. 

The paper is structured as follows. In Sec.~\ref{secII} we present the exact solutions for the Higgs sector of the Standard Model and the the Green functions of the classical theory. In Sec.~\ref{secIIb} we present the solutions of the Dyson-Schwinger equations till the 2-point functions. In Sec.~\ref{secIII} we discuss the gap equations we obtained. Finally, in Sec.~\ref{secIV} we yield the conclusions.

\section{Exact solutions \label{secII}}

\subsection{Higgs sector}

Let us consider the Higgs sector of the Standard Model given by \cite{Quigg:2013ufa}
\begin{equation}
    {\cal L}_H=\partial_\mu\varphi^\dagger\partial^\mu\varphi-\mu^2|\varphi|^2-\lambda|\varphi|^4
\end{equation}
being
\begin{equation}
   \varphi =\left(
	 \begin{array}{c}
		 \phi^{+} \\
		 \phi^{0}
	 \end{array}
	\right)
\end{equation}
and $|\varphi|^2=|\phi^{+}|^2+|\phi^{0}|^2$. Then, the equations of motion are
\begin{eqnarray}
\label{eq:eom}
     \partial^2\phi^{+}&=&-\mu^2\phi^{+}-2\lambda(|\phi^{+}|^2+|\phi^{0}|^2)\phi^{+} \\
		 \partial^2\phi^{0}&=&-\mu^2\phi^{0}-2\lambda(|\phi^{+}|^2+|\phi^{0}|^2)\phi^{0}.
\end{eqnarray}
Here and below we take $v^2=|\mu^2|/(2\lambda)$ for the vacuum expectation value of the theory,
as usual in the Higgs mechanism. 

In the following we will limit our analysis just to this case with no interaction with gauge bosons and fermion fields. The reason to use this approximation is that, for such case, exact solutions to the quantum field theory can be found as we see below. 

\subsection{Classical solutions\label{subs:cs}}

We consiser two different phases, U(nbroken) and B(roken), and to solve the eq.(\ref{eq:eom}), we write
\begin{eqnarray}
\label{eq:exsols}
    \phi^{+}(x)&=&e^{i\xi_+}h_{U}(x) \\
		\phi^0(x)&=&e^{i\xi_0}h_{U,B}(x).
\end{eqnarray}
In both cases we obtain two solutions. For the unbroken phase, $\mu^2>0$, one has either the trivial solution $h_U(x)=0$ or the non-trivial one given by
\begin{equation}
   h_U(x) = \Lambda\left(\frac{1}{3\lambda}\right)^\frac{1}{4}{\rm sn}\left(p\cdot x+\theta_U,-\frac{\Lambda^2\sqrt{3\lambda}}{2\mu^2+\Lambda^2\sqrt{3\lambda}}\right),
\end{equation}
provided the following dispersion relation does hold
\begin{equation}
\label{eq:drun}
    p^2=\mu^2+\frac{\Lambda^2}{2}\sqrt{3\lambda}.
\end{equation}
Here $\Lambda$ and $\theta_U$ being two arbitrary integration constants and sn a Jacobi elliptic function with parameter $k^2=-\frac{\Lambda^2\sqrt{3\lambda}}{2\mu^2+\Lambda^2\sqrt{3\lambda}}$. 
From these we realize that, already at classical level, the mass $\mu$ is modified by the self-interaction with a coupling $\lambda$. We just need a finite self-coupling to get such a solution. This solution reduces to the known case for $\mu=0$ given in \cite{Frasca:2009bc,Frasca:2013tma}.

For B solution one has (we assume the field $\phi^+$ to be zero) either the Standard Model solution $h_B(x)=v$ that is stable (the other one, $h_B(x)=0$, is unstable) or
\begin{equation}
\label{eq:solnt}
   h_B(x) = \left(\frac{|\mu|^2}{3\lambda}\right)^\frac{1}{2}\ {\rm dn}\left(p\cdot x+\theta_B,-1\right)
\end{equation}
being dn another Jacobi elliptic function of parameter $k^2=-1$, with $\theta_B$ and $\Lambda$ the energy scale both arbitrary integration constants. This holds given the following dispersion relation
\begin{equation}
\label{eq:drB}
   p^2=\frac{|\mu|^2}{3}.
\end{equation}

\subsection{Green functions}

In order to get the Green functions of the classical theory, we start from the equations of motion given by
\begin{eqnarray}
\label{eq:eom2}
     \partial^2\phi^{+}&=&-\mu^2\phi^{+}-2\lambda(|\phi^{0}|^2+|\phi^{+}|^2)\phi^{+}+j_+ \\
		 \partial^2\phi^{0}&=&-\mu^2\phi^{0}-2\lambda(|\phi^{0}|^2+|\phi^{+}|^2)\phi^{0}+j_0 \\
     \partial^2\phi^{+*}&=&-\mu^2\phi^{+*}-2\lambda(|\phi^{0}|^2+|\phi^{+}|^2)\phi^{+*}+j_+^* \\
		 \partial^2\phi^{0*}&=&-\mu^2\phi^{0*}-2\lambda|(\phi^{0}|^2+|\phi^{+}|^2)\phi^{0*}+j_0^*
\end{eqnarray}
with respect to $j_+$ and $j_0$ and $j_+^*$ and $j_0^*$. We can see that the equations for the Green functions have no nonlinear term. The only physical component for the Higgs field is the real component of $\phi^{0}$. 
We consider $\phi_0$ as a functional of $j_0$. Then, the classical Green functions are given by the conditions
\begin{equation}
\label{eq:Gcon}
      \left.\frac{\delta\phi^{0*}}{\delta j_0(y)}\right|_{j_0,j_{0}^*=0}=
			\left.\frac{\delta\phi^{0}}{\delta j_{0}^*(y)}\right|_{j_0,j_{0}^*=0}=0.
\end{equation}
These yield the general result
\begin{equation}
     \partial^2\Delta_H(x,y)-|\mu|^2\Delta_H(x,y)+6\lambda[\varphi_B(x)]^2\Delta_H(x,y)=\delta^4(x-y).
\end{equation}
Firstly, we consider the Standard Model solution
\begin{equation}
\phi^+=0, \qquad \phi_0=v,  
\end{equation}
being $v$ a constant. From eq.(\ref{eq:Gcon}) we will get
\begin{equation}
\label{eq:GSM}
     \partial^2\Delta_{SM}(x,y)+2|\mu^2|\Delta_{SM}(x,y)=\delta^4(x-y).
\end{equation}
Then, the classical Green function takes the simple form in momentum space
\begin{equation}
    \Delta_{SM}(x,y)=\frac{1}{p^2-2|\mu^2|+i\epsilon}
\end{equation}
where we have introduced the $i\epsilon$ prescription for the Fourier transform.
The other solution is more involved. One has \cite{Frasca:2015wva}
\begin{equation}
\label{eq:prop}
     \Delta_H(p)=\frac{\sqrt{2}\pi^3}{K^3(-1)}\sum_{n=1}^\infty n^2\frac{e^{-n\pi}}{1+e^{-2n\pi}}\frac{1}{p^2-m_n^2+i\epsilon}
\end{equation}
being
\begin{equation}
\label{eq:spe}
    m_n=n\frac{\pi}{K(-1)}\frac{|\mu|}{\sqrt{3}}
\end{equation}
the mass spectrum that also entails a zero mass value, the Goldstone boson. $K(-1)$ is a complete elliptic integral of the first kind. We see that the spectrum has a Kaluza-Klein form in this case.

Given eqs.(\ref{eq:GSM}) and (\ref{eq:prop}), the classical theory for both the solutions is completely solved as already shown in~\cite{Frasca:2013tma}. At this stage we dub Standard Model solution the one found in textbooks and the background field solution the one we obtained.

\section{Dyson-Schwinger equations \label{secIIb}}

The background field solution to the hierachy of Dyson-Schwinger equations for the Higgs field was firtsly presented in \cite{Frasca:2015wva}. It is based on a technique devised in \cite{Bender:1999ek}. We apply it to the only surviving component of the Higgs field that now we name $\phi_H$.

The generating functional for this case is given by
\begin{equation}
    Z[j]=Z_0\int[d\phi]e^{i\int d^4x\left[\frac{1}{2}(\partial\phi_H)^2+\frac{1}{2}|\mu^2|\phi_H^2-\frac{2\lambda}{4}\phi_H^4+j\phi_H\right]}.
\end{equation}
Our aim is to evaluate the 1-point and 2-point correlation functions. We take the average on the vacuum state $|0\rangle$ and divide by $Z[j]$ giving
\begin{equation}
\label{eq:sf1}
    \partial^2 G_1^{(j)}(x)-|\mu^2|G_1^{(j)}(x)+2\lambda\frac{\langle 0|\phi_H^3|0\rangle}{Z[j]}=j
\end{equation}
having defined $G_1^{(j)}(x)=\langle 0|\phi_H|0\rangle/Z[j]$, the one-point function. We take
\begin{equation}
    G_1^{(j)}(x)Z[j]=\langle 0|\phi|0\rangle
\end{equation}
and we compute the functional derivative with respect to $j$ giving
\begin{equation}
    [G_1^{(j)}(x)]^2Z[j]+G_2^{(j)}(x,x)Z[j]=\langle 0|\phi^2|0\rangle
\end{equation}
and, deriving once again, one has
\begin{equation}
    [G_1^{(j)}(x)]^3Z[j]+3G_2^{(j)}(x,x)G_1^{(j)}(x)Z[j]+G_3^{(j)}(x,x,x)Z[j]=\langle 0|\phi^3|0\rangle.
\end{equation}
Using eq.(\ref{eq:sf1}), this becomes
\begin{equation}
\label{eq:g1}
   \partial^2 G_1^{(j)}(x)-|\mu^2|G_1^{(j)}(x)+2\lambda\left([G_1^{(j)}(x)]^3+3G_2^{(j)}(x,x)G_1^{(j)}(x)+G_3^{(j)}(x,x,x)\right)=j.
\end{equation}
Taking $j=0$, observing that the theory is invariant by translations, that is $G_2(x,y)=G_2(x-y)$, one has the first Dyson-Schwinger equation of the scalar theory
\begin{equation}
\label{eq:g10}
   \partial^2 G_1(x)-|\mu^2|G_1(x)+2\lambda\left([G_1(x)]^3+3G_2(0)G_1(x)+G_3(0,0)\right)=0.
\end{equation}
Now, we absorb the renormalization constant $G_2(0)$ into the mass as $\mu_R^2=|\mu^2|-6\lambda G_2(0)$. 
This yields
\begin{equation}
\label{eq:g101}
   \partial^2 G_1(x)-\mu_R^2G_1(x)+2\lambda[G_1(x)]^3=0.
\end{equation}
This equation admits exact solutions, respecting translation invariance. Before to see this, we derive the Dyson-Schwinger equation for the two-point function. We take the functional derivative of eq.(\ref{eq:g1}) to obtain
\begin{eqnarray}
   &&\partial^2G_2^{(j)}(x,y)-|\mu^2|G_2^{(j)}(x,y)+2\lambda\left(3[G_1^{(j)}(x)]^2G_2^{(j)}(x,y)+3G_2^{(j)}(x,x)G_2^{(j)}(x,y)\right. \nonumber \\
	&&\left.+3G_3^{(j)}(x,x,y)G_1^{(j)}(x)+G_4^{(j)}(x,x,x,y)\right)=\delta^4(x-y).
\end{eqnarray}
We substitute $j=0$ into this equation to obtain
\begin{eqnarray}
\label{eq:g20}
   &&\partial^2G_2(x-y)-\mu_R^2G_2(x-y)+2\lambda\left(3[G_1(x)]^2G_2(x-y)\right. \nonumber \\
	&&\left.+3G_3(0,x-y)G_1(x)+G_4(0,0,x-y)\right)=\delta^4(x-y).
\end{eqnarray}
Again, we set $G_3$ and $G_4$ to zero. Then,
\begin{equation}
\label{eq:g201}
   \partial^2G_2(x-y)-\mu_R^2G_2(x-y)+6\lambda[G_1(x)]^2G_2(x-y)=\delta^4(x-y).
\end{equation}
The solutions to eq.(\ref{eq:g101}) are $G_1(x)=0$, unstable, $G_1(x)=v$, stable, and
\begin{equation}
\label{eq:bst}
    G_1(x)=\left(\frac{\mu_R^2}{3\lambda}\right)^\frac{1}{2}\ {\rm dn}\left(p\cdot x+\chi_B,-1\right)
\end{equation}
being $\chi_B$ an arbitrary integration constant and provided we put $G_3(0,0)=0$,
that is a consistent choice as we will show below,
and choosing the momenta $p$ to satisfy
\begin{equation}
   p^2=\frac{\mu_R^2}{3}.
\end{equation}
Two-point functions, both for the Standard Model case and the background field solution, are given by eqs.(\ref{eq:GSM}) and (\ref{eq:prop}) provided we substitute $|\mu|^2\rightarrow \mu^2_R$ everywhere. We just note that the equations for $G_1$ and $G_2$ apply for both the solutions

\section{Gap equations \label{secIII}}

We discuss both the cases for the Standard Model solution and the background field solution (\ref{eq:bst}). The former corresponds to the current approach used in the Standard Model and presently supported by experimental data obtained at LHC so far. The latter yields a possible view of the Higgs boson as a Kaluza-Klein excitation and emerges naturally as an exact solution to the Dyson-Schwinger hierarchy of equations for the sole Higgs sector of the Standard Model.

\subsection{Standard Model solution}

The equation for the Green function of the Higgs field is given by eq.(\ref{eq:g201}), provided we take $G_1(x-y)=v$. This yields
\begin{equation}
\label{eq:g220}
  \partial^2G_2(x-y)-\mu_R^2G_2(x-y)+6\lambda[G_1(x-y)]^2G_2(x-y)=\delta^4(x-y),
\end{equation}
remembering that $\mu^2_R=|\mu^2|-6\lambda G_2(0)$.Then,
\begin{equation}
\label{eq:g221}
  \partial^2G_2(x-y)+m_H^2G_2(x-y)=\delta^4(x-y),
\end{equation}
being $-m_H^2=2|\mu^2|-6\lambda G_2(0)$, the renormalized Higgs mass. This yield the self-consistent equation for the Higgs mass
\begin{equation}
-m_H^2=2|\mu^2|-6\lambda\int\frac{d^4p}{(2\pi)^4}\frac{1}{p^2-m_H^2+i\epsilon}.
\end{equation}
The integral can be exactly evaluated to give
\begin{equation}
m_H^2(|\mu|,\lambda,\Lambda)=-2|\mu^2|+\frac{3\lambda}{16\pi^2}\left[\Lambda^2-m_H^2(|\mu|,\lambda,\Lambda)\ln\left(1+\frac{\Lambda^2}{m_H^2(|\mu|,\lambda,\Lambda)}\right)\right]
\end{equation}
If we choose, $|\mu^2|=2\lambda v^2$, $\lambda=0.1$, $v=246\ GeV$, when $\Lambda\approx 1695\ GeV$ we get the correct experimental mass of the Higgs of about 125 GeV. Otherwise, one can consider the Higgs mass as a function of three independent variables, $v,\ \lambda$ and $\Lambda$, and get always a finite value for the Higgs mass. This yields a running coupling, fixed $v$ and $m_H$ to their experimental values,
\begin{equation}
\lambda(\Lambda)=\frac{m_H^2}{-4v^2+\frac{3}{16\pi^2}\left[\Lambda^2-m_H^2\ln\left(1+\frac{\Lambda^2}{m_H^2}\right)\right]}.
\end{equation}
We just point out that this does not solve the naturalness problem but gives a correct value of the Higgs mass anyway, provided one fixes $\Lambda$ to some given values.

\subsection{Background field solution}

We consider the 2-point correlation function given by eq.(\ref{eq:prop}). This holds given a spectrum of the theory in the form
\begin{equation}
m_n=n\frac{\pi}{\sqrt{3}K(-1)}\left||\mu|^2-6\lambda\int\frac{d^4p}{(2\pi)^4}\Delta_H(p)\right|^\frac{1}{2},
\end{equation}
where $\Delta_H(p)$, is taken again from eq.(\ref{eq:prop}). This is a self-consistent equation to determine $m_n$ and can be cast in an analytical form by evaluating the integral. Now, we can set
\begin{equation}
m_n=nm_H
\end{equation}
yielding
\begin{equation}
m_H=\frac{\pi}{\sqrt{3}K(-1)}\left||\mu|^2-6\lambda\sum_{n=1}^\infty B_n\int\frac{d^4p}{(2\pi)^4}\frac{1}{p^2-n^2m_H^2+i\epsilon}\right|^\frac{1}{2}
\end{equation}
being
\begin{equation}
B_n=\sqrt{2}\frac{\pi^3}{K^3(-1)}n^2\frac{e^{-n\pi}}{1+e^{-2n\pi}}.
\end{equation}
We can evaluate the integral and take the square obtaining
\begin{equation}
m_H^2(|\mu|,\lambda,\Lambda)=\frac{\pi^2}{3K^2(-1)}\left\{-|\mu|^2+\frac{3\lambda}{16\pi^2}\left[\Lambda^2-\sum_{n=1}^\infty B_nn^2m_H^2(|\mu|,\lambda,\Lambda)
\ln\left(1+\frac{\Lambda^2}{n^2m_H^2(|\mu|,\lambda,\Lambda)}\right)\right]\right\}.
\end{equation}
where use has been made of the fact that $\sum_{n=1}^\infty B_n=1$. 
We can yield a running coupling at a fixed Higgs mass $m_{\rm higgs}$, using $|\mu|^2=2\lambda v^2$, as
\begin{equation}
\lambda(\Lambda)=\frac{m_{\rm higgs}^2}{\frac{\pi^2}{3K^2(-1)}\left\{-2v^2+\frac{3}{16\pi^2}\left[\Lambda^2
-m_{\rm higgs}^2\sum_{n=1}^\infty B_nn^2\ln\left(1+\frac{\Lambda^2}{n^2m_{\rm higgs}^2}\right)\right]\right\}}.
\end{equation}
It is easy to get the $\beta$ function from here giving
\begin{equation}
\beta(\lambda)=\frac{1}{16\pi^2K^2(-1)}\left(1-\sum_{n=1}^\infty B_nn^2\frac{n^2}{n^2+1}\right)\lambda^2
\end{equation}
that is very much similar to the textbook result \cite{ZinnJustin:2002ru}. 

We emphasize that a spectrum like the one obtained in this case, besides having a Kaluza-Klein form, implies that at least a resonance with the double of the mass of the Higgs particle, about 250 GeV, should be seen at LHC. Other higher excitations are strongly damped and so, very difficult to observe. We are consistent with the theoretical expectations also given in \cite{Hill:2019cce}.
 
\section{Conclusions \label{secIV}}

We have seen how the Higgs sector without interactions can be solved exactly. Two classes of solutions emerge: One is the Higgs field with a satisfactory wealth of data from LHC and the other one is a solution with a background field manifesting a Kaluza-Klein spectrum.

Production processes observed at LHC do not exclude the possibility of higher excited states. Anyway, it appears that the simplest solution seems the one nature selected so far.

Eventually, observation of the effects of the background field solution at LHC should be accomplished in the study of the signal of the 4-lepton decay of the Higgs via the ZZ process. The reason for this is the difficulty to observe higher excited states due to their very low probability of production. It should be seen a possible excitation at the double of the mass of the Higgs particle, about 250 GeV. As we have seen, renormalization effect can slightly change this value.

It is rather interesting that the Higgs sector could possess a non-trivial solution with a spectrum of a Kaluza-Klein particle that, if confirmed, could be a clue of existence of a further space dimension.




\end{document}